\documentstyle[psfig,cite]{article}

\def\simgt{\rlap{\lower 3.5 pt \hbox{$\mathchar \sim$}}%
           \raise 1pt \hbox {$>$}}
\def\simlt{\rlap{\lower 3.5 pt \hbox{$\mathchar \sim$}}%
           \raise 1pt \hbox {$<$}}

\def\gev{{\rm GeV}}
\def\tev{{\rm TeV}}
\def\mt{m_t^{}}
\def\mh{m_H^{}}

\def\mh{m_H^{}}
\def\mz{m_Z^{}}
\def\mw{m_W^{}}
\def\ov{\overline}
\def\msbar{\ov{\rm MS}}
\def\sm{\rm SM}

\def\delg{\bar{\delta}_G^{}}
\def\delgsm{\bar{\delta}_G^{\sm}}
\def\ewgroup{{\rm SU(2)_L \times U(1)_Y}}

\newcommand{\beq}{\begin{equation}}
\newcommand{\eeq}{\end{equation}}
\newcommand{\bea}{\begin{eqnarray}}
\newcommand{\eea}{\end{eqnarray}}
\newcommand{\bsub}{\begin{subequations}}
\newcommand{\esub}{\end{subequations}}

\makeatletter
\newtoks\@stequation

\def\subequations{\refstepcounter{equation}%
  \edef\@savedequation{\the\c@equation}%
  \@stequation=\expandafter{\theequation}
  \edef\@savedtheequation{\the\@stequation}
  \edef\oldtheequation{\theequation}%
  \setcounter{equation}{0}%
  \def\theequation{\oldtheequation\alph{equation}}}

\def\endsubequations{%
  \ifnum\c@equation < 2 \@warning{Only \the\c@equation\space subequation
    used in equation \@savedequation}\fi
  \setcounter{equation}{\@savedequation}%
  \@stequation=\expandafter{\@savedtheequation}%
  \edef\theequation{\the\@stequation}%
  \global\@ignoretrue}

\def\eqnarray{\stepcounter{equation}\let\@currentlabel\theequation
\global\@eqnswtrue\m@th
\global\@eqcnt\z@\tabskip\@centering\let\\\@eqncr
$$\halign to\displaywidth\bgroup\@eqnsel\hskip\@centering
     $\displaystyle\tabskip\z@{##}$&\global\@eqcnt\@ne
      \hfil$\;{##}\;$\hfil
     &\global\@eqcnt\tw@ $\displaystyle\tabskip\z@{##}$\hfil
   \tabskip\@centering&\llap{##}\tabskip\z@\cr}

\makeatother

\begin{document}
\thispagestyle{empty}
\vspace*{-15mm}
\baselineskip 10pt
\begin{flushright}
\begin{tabular}{l}
{\bf CERN-TH/97-349}\\
{\bf KEK-TH-550}\\
{\bf hep-ph/9712260}
\end{tabular}
\end{flushright}
\baselineskip 18pt 
\vglue 15mm 

\begin{center}
{\Large\bf
Constraints on Non-Standard Contributions to the Charged-Current Interactions
}
\vspace{5mm}

\def\thefootnote{\alph{footnote}}
\setcounter{footnote}{0}
{\bf
Kaoru Hagiwara$^{1,2}$ and 
Seiji Matsumoto$^{3}$
}
\vspace{5mm}

$^1${\it Theory Group, KEK, Tsukuba, Ibaraki 305, Japan}\\
$^2${\it ICEPP, University of Tokyo, Hongo, Bunkyo-ku, Tokyo 113, Japan}\\
$^3${\it TH Division, CERN, CH-1211 Geneva 23, Switzerland}
\end{center}

\vspace{15mm}

\begin{center}
{\bf Abstract}\\[10mm]
\begin{minipage}{12cm}
\noindent
The success of the quantum level predictions of the Standard Model on 
the $Z$ boson properties, on $\mw$ and on $\mt$,  
which makes use of the muon lifetime as an input, 
implies a stringent constraint on new physics contributions to the 
$V-A$ charged-current interactions among leptons. 
Observed unitarity of the CKM matrix elements 
then implies constraints on 
non-standard contributions to the lepton-quark charged-current 
interactions. 
By using the recent 
electroweak data as inputs, we find the 95\%~CL 
limits for the corresponding contact interactions:  
$\Lambda_{CC,+}^{\ell\ell}>7.5$~TeV and 
$\Lambda_{CC,-}^{\ell\ell}>10.2$~TeV 
for the purely leptonic terms, and 
$\Lambda_{CC,+}^{\ell q}>5.8$~TeV and 
$\Lambda_{CC,-}^{\ell q}>10.1$~TeV 
for the lepton-quark contact interactions.  
\end{minipage}
\end{center}

\vspace{20mm}
\noindent
\begin{tabular}{l}
{\bf CERN-TH/97-349}\\
December 1997
\end{tabular}

\thispagestyle{empty}
\vfil
\newpage

The reports from HERA~\cite{h197,zeus97} of event rates above Standard Model 
(SM) expectations in $e^+p\to e^+X$ deep inelastic scattering at very high 
$Q^2$ renewed the interest in the present constraints on non-standard 
interactions 
among leptons and quarks~\cite{altarelli97,babu97,bchz97,contact97}.  
In particular, studies on low energy neutral current experiments find 
stringent constraints on parity-violating lepton-quark interactions~\cite%
{altarelli97,babu97,bchz97,contact97,chm97}, and those including high-$Q^2$ 
experiments at the Tevatron and LEP2 find constraints on general lepton-quark 
contact interactions in the neutral currents~\cite{bchz97,bchz97b}.   
In terms of the ``compositeness scale''~\cite{elp82,pdg96}, 
the 95\%~CL lower bounds are found to be typically of the order of 
10~TeV for chiral contact terms while they can be as low as 3~TeV for 
those contact terms that conserve parity or with purely axial-vector 
quark currents~\cite{bchz97b}.
In view of the above constraints, it is unlikely that the observed 
excess of high-$Q^2$ events at HERA can be explained as a consequence 
of new chirality and flavour-conserving neutral currents between leptons 
and quarks.

As for charged-current (CC) $e^+p\to \bar{\nu}_e X$ events, both H1 and ZEUS 
experiments observed about a factor of 2 more events than expected 
in the SM at $Q^2>10,000\,\gev^2$~\cite{h197,heracc97}.  
Although the statistics of high-$Q^2$ events is still limited, 
one may hope to probe new interactions at HERA also in the charged 
currents~\cite{rueckl83,burges84,cashmore85,burgess94,%
chiappetta96,cornet97,altarelli}.  

In this letter, we would like to point out that non-standard contributions  
to the CC interactions are strongly constrained by the 
electroweak measurements, which are consistent with the predictions of 
the minimal SM and also by the observed unitarity of the Cabibbo-Kobayashi-%
Maskawa (CKM) matrix elements.  
Normally, one takes as inputs 
the observed muon decay constant $G_F$, the $Z$ boson 
and the top-quark 
masses, and an estimate of the running QED coupling 
$\alpha(m_Z^2)$. The predictions are made for all the other electroweak 
observables, including various asymmetries on the $Z$ pole and the $W$ mass, 
as functions of an assumed Higgs boson mass $\mh$.  
We show that by using all the electroweak data as inputs besides $G_F$ 
we can calculate the muon decay constant in the minimal SM, and 
that by comparing it with the observed value we can obtain constraints 
on non-SM contributions to the leptonic charged currents.  
The observed unitarity of the CKM matrix elements then constrains the 
difference between new physics contributions to the purely leptonic and 
lepton-quark CC interactions.  

Effects of new physics at energies below new particle thresholds 
can generally be expressed as non-renormalizable higher-dimensional 
terms in the effective Lagrangian of the SM particles.  
The effective CC interactions among quarks and leptons 
may be parametrized as 
\beq
{\cal L}_{CC} = \sum_{f_i} \sum_{\alpha, \beta}\ 
\eta^{f_1 f_2 f_3 f_4}_{\alpha \beta}\  S
\ov{\psi_{f_1}} \gamma^\mu P_\alpha \psi_{f_2}\ 
\ov{\psi_{f_3}} \gamma_\mu P_\beta \psi_{f_4}\,, 
\label{etadef}
\eeq
where $(f_1,f_2)$ and $(f_3,f_4)$ are lepton and/or quark charged currents, 
$\alpha,\beta = L,R$ denote their chirality: 
$P_{L} = (1-\gamma_5)/2$, 
$P_{R} = (1+\gamma_5)/2$, 
and $S$ is a statistical factor, which is $1/2$ if the two currents are 
the same and is otherwise 1. 
The coefficients $\eta^{f_1 f_2 f_3 f_4}_{\alpha \beta}$ have the 
dimension of (mass)$^{-2}$, which may be expressed as~\cite{elp82,pdg96}
\beq
\eta^{f_1 f_2 f_3 f_4}_{\alpha \beta} = \epsilon \frac{4\pi}%
{(\Lambda^{f_1 f_2 f_3 f_4}_{\alpha \beta, \epsilon })^2}\,,
\label{lambdadef} 
\eeq
with the sign factor $\epsilon=+$ or $-$.
If the contact interactions are results of an exchange of 
an extra heavy charged vector boson $W_E$, they are given by 
\beq
\eta^{f_1 f_2 f_3 f_4}_{\alpha \beta} = 
-\frac{g_\alpha^{f_1 f_2} g_\beta^{f_3 f_4}}{m_{W_E}^2},
\label{extraw}
\eeq
where $g_\alpha^{f_1 f_2}$ and $g_\beta^{f_3 f_4}$ are the $W_E$-boson 
couplings to the corresponding lepton and quark currents. 

In this report we study only the $V-A$ contact interactions, 
$\alpha=\beta=L$ terms in eq.~(\ref{etadef}), 
because they interfere with the SM amplitudes and 
because, without light right-handed neutrinos, $V+A$ currents do not 
contribute to low energy leptonic or lepton-quark CC processes.   
Also, we do not consider mixing between the SM $W$ boson and possible 
additional $W_E$ bosons.  
The latter assumption allows us to use the observed $W$ mass in the 
SM contributions to the muon decay amplitude.   

Then, the observed Fermi constant $G_F$ is expressed as 
\bea
4\sqrt{2}G_F &=& \frac{\bar{g}_W^2(0) +\hat{g}^2\delg}{m_W^2} 
-\eta_{LL}^{e \nu_e \nu_\mu \mu}
\nonumber \\
&\approx& \frac{\bar{g}_W^2(0)}{m_W^2}\big[ 1 +\delg 
-\frac{\eta_{LL}^{e \nu_e \nu_\mu \mu}}{4\sqrt{2}G_F} \big]\,.
\label{gf_with_eta}
\eea
Here $\bar{g}_W^2(0)$ and $\delg$ are the universal charge form factor
and the radiative corrections coming from vertex and box diagrams 
to the $\mu$-decay, respectively.  
The hatted symbols, $\hat{s}$ and $\hat{g}$, denote 
$\msbar$ couplings (see Ref.~\cite{hhkm94} for the notation). 
{}From eq.~(\ref{gf_with_eta}), we see immediately that
the effect from the new contact term 
$\eta_{LL}^{e \nu_e \nu_\mu \mu}$ 
can be regarded as a 
non-standard contribution to the parameter 
$\delg$. 
In Ref.~\cite{hhkm94}, it has been shown that only 
a combination $T'$ such that
\begin{eqnarray}
  T' &=& T +\frac{\delgsm -\delg}{\alpha}
\end{eqnarray}
is constrained by the electroweak experiments
in the presence of non-standard contributions to $\delg$. 
Here the parameter $T$~\cite{stu} is a measure of the difference between the 
neutral- and charged-current interactions at low energies.  
The effect of the new contact interaction can hence 
be constrained by the difference $\Delta T$ between 
the measured $T'$ parameter and the SM prediction 
for the $T$ parameter:
\begin{eqnarray}
   \Delta T = T' -T_{\rm SM}
   &=& \frac{\eta_{LL}^{e \nu_e \nu_\mu \mu}}{4\sqrt{2}G_F\,\alpha }\,.
\end{eqnarray}

In the following analysis we use recent electroweak data%
\footnote{
We do not include the low energy neutral current (LENC) data 
(see Ref.~\cite{chm97}) in this letter 
because they may be affected by the presence of new 
contact interactions in the neutral current sectors. 
The properties of the $Z$ and $W$ bosons 
on the pole are not affected by the new interactions 
as long as the mixing between the SM weak bosons 
and the new vector bosons are negligible. 
}
of the $Z$ parameter measurements at LEP1 and SLC~\cite{ew97}
and the $W$ mass measurements at the Tevatron and LEP2~\cite{ew97}.
In addition, we use the combined value of the top-quark mass 
measured at CDF and D0, 
$m_t=175.6\pm 5.5\,\gev$~\cite{ew97}, 
the QCD coupling estimated by PDG,
$\alpha_s=0.118\pm 0.003$~\cite{pdg96}, and 
the estimate of the QED coupling at the $\mz$ scale, 
$\delta_\alpha \equiv 1/\bar{\alpha}(m_Z^2)-128.72
=0.03\pm 0.09$~\cite{eidelman95}. 
We calculate all radiative corrections within 
the SM, except that we allow a new contribution $\Delta T$ to 
the $T'$ parameter 
(see Refs.~\cite{hhkm94,hhm97} for details).  

{}From the fit to the $Z$ boson parameters and the $W$ mass, 
we obtain
\bea
\Delta T 
\equiv T' -T_{\rm SM}
= -0.06^{+0.26}_{-0.11} \:\:\: {\rm or} \:\:\: 
 \eta_{LL}^{e \nu_e \nu_\mu \mu}
 = (-0.03^{+0.13}_{-0.05})\tev^{-2}\,.
 \label{delta_t}
\eea
as the 1$\sigma$ constraints.
In Fig.1, we show 
the minimal $\chi^2$ as a function of $\Delta T$, where the Higgs boson 
mass $\mh$ is allowed to vary in the fit.  
The solid line is obtained by constraining $\mh$ in the 
range $77.1\, \gev <\mh<1\,\tev$:  
the lower limit is from direct measurements at LEP2~\cite{mhlimit97} while   
our perturbative calculation is unreliable at $\mh\, \simgt\, 1\,\tev$.  

For comparison, we also give the result obtained 
by allowing $\mh$ to vary freely, without the boundaries. 
As seen from the figure, both results are
the same for $-0.1\,\simlt\, \Delta T\, \simlt\, 0.35$ 
because the fitted $\mh$ value stays 
in the allowed region. 
On the other hand, for $\Delta T\,\simlt -0.1$, 
the best fit value of $\mh$ is found below $77.1\,\gev$
for the dashed line.
This explains the difference between the two curves. 
The constraint (\ref{delta_t}) and all the following results are obtained 
by requiring $77.1\,\gev<\mh<1\,\tev$. 

In order to obtain the 95\% confidence level limit 
for $\eta$, we assume that the probability density 
function $P(\eta)$ is proportional to $\exp(-\chi^2(\eta)/2)$. 
Further we need to consider two cases separately to 
convert the limit for $\eta$ to that for $\Lambda$~\cite{bchz97b}. 
The 95\% CL limit on $\Lambda_+$, which corresponds to 
$\epsilon=+1$ in eq.~(\ref{lambdadef}), is obtained from 
$\eta_+=4\pi/\Lambda_+^2$, where
 \begin{eqnarray}
    0.05 = \frac{\int_{\eta_+}^{\infty}P(\eta)d\eta}
                {\int_{0}^{\infty}P(\eta)d\eta} \,.
    \label{probabiblity1_norm}
 \end{eqnarray}
Likewise
the 95\% CL limit on $\Lambda_-$ is obtained 
from $\eta_-=-4\pi/\Lambda_-^2$, where
 \begin{eqnarray}
  0.05 = \frac{\int^{\eta_-}_{-\infty}P(\eta)d\eta}
                {\int^{0}_{-\infty}P(\eta)d\eta} \,.
    \label{probabiblity2_norm}
 \end{eqnarray}
We find the 95\%CL limits for $\Lambda_{\pm}$
\footnote{
In order to test the sensitivity of the above bounds to the 
estimate of $\bar{\alpha}(m_Z^2)$, we repeat the analysis by 
using the estimate $\delta_\alpha=0.12\pm 0.06~$\cite{mz95}.  
We find 
$\Lambda_+>7.8\,\tev$ and $\Lambda_->9.7\,\tev$
as the 95\% CL limits, which do not differ much 
from the results (\ref{lambda95}) that are obtained with the estimate 
$\delta_\alpha=0.03\pm 0.09~$\cite{eidelman95}.  
}: 
\bsub \label{lambda95}
\bea
\Lambda_{LL,+}^{e \nu_e \nu_\mu \mu} \equiv \Lambda_{CC,+}^{ll} 
&>& 7.5\,\tev  \,,\\
\Lambda_{LL,-}^{e \nu_e \nu_\mu \mu} \equiv \Lambda_{CC,-}^{ll} 
&>& 10.2\,\tev \,. 
\eea
\esub

Next we consider the limit on 
the lepton-quark contact interactions. 
We assume the unitarity of the CKM in the SM:
 \begin{eqnarray}
    |V_{ud}^{\rm SM}|^2 +|V_{us}^{\rm SM}|^2 +|V_{ub}^{\rm SM}|^2 = 1 \,.
    \label{unitarity_sm}
 \end{eqnarray}
Experimentally, this unitarity is slightly violated at 
the 1.7$\sigma$ level~\cite{pdg96}:
\bea
|V_{ud}|^2 +|V_{us}|^2 +|V_{ub}|^2 = 0.9965 \pm 0.0021\,.
 \label{unitarity_exp}
\eea
In the presence of new contact interactions in the CC
sector, the observed CKM matrix elements can be 
shifted as 
\begin{eqnarray}
  V_{u d_j}^{\rm obs} &=& V_{u d_j}^{\sm}
 -\frac{\eta_{LL}^{\ell \nu_\ell u d_j} 
         -V_{u d_j}^{\rm SM}\eta_{LL}^{e \nu_e \nu_\mu \mu}}
       {4\sqrt{2}G_F}
  \label{v_obs}
\end{eqnarray}
because the CKM matrix elements are experimentally determined 
from the ratio of the semileptonic and the purely leptonic
CC strengths
\footnote{
By setting $\eta^{e\nu_e \nu_\mu \mu}_{LL} = 0$, 
the formula (\ref{v_obs}) reduces to that of Altarelli et.al.~\cite{altarelli}
who neglect contact interactions to the purely leptonic 
channel.
}.

In general, three lepton-quark contact interactions 
in the $V-A$ charged current, 
$\eta^{\ell\nu_\ell ud}_{LL}$,
$\eta^{\ell\nu_\ell us}_{LL}$,
$\eta^{\ell\nu_\ell ub}_{LL}$,
can enter the unitarity constraint (\ref{unitarity_exp}).
By assuming e.g. 
$\eta^{\ell\nu_\ell us}_{LL}=\eta^{\ell\nu_\ell ub}_{LL}=0$,
one can obtain constraint on
$\eta^{\ell\nu_\ell ud}_{LL}- V_{ud}\eta^{e\nu_e \nu_\mu \mu}_{LL}$.
In this paper, we examine the case where the 
contact interactions satisfy the 
$\ewgroup$ gauge symmetry of the SM.  
Then $V-A$ charged-current interactions among leptons and those 
between leptons and 
quarks are expressed in terms of left-handed lepton doublets, 
$L_\ell=(\nu_\ell,\ell_L)$ $(\ell=e,\mu,\tau)$, and quark doublets
$Q_i=(u_{iL},d_{iL})$ ($i=1,2,3$).  
Assuming that the new interactions do not distinguish 
between quark and lepton generations, 
CC contact interactions should have
the following universal form~\cite{bchz97b} 
\bea
{\cal L}_{\rm SU(2)} &=&
\frac{\eta_2^{LL}}{1+\delta_{ij}} 
\Bigl(\ov{L_i}\gamma^\mu T^a L_i\Bigr) \!
\Bigl(\ov{L_j}\gamma_\mu T^a L_j\Bigr) 
\!+\eta_2^{LQ} 
\Bigl(\ov{L_i}\gamma^\mu T^a L_i\Bigr) \!
\Bigl(\ov{Q_j}\gamma_\mu T^a Q_j\Bigr) .\quad\,\,\,
\label{lsu2}
\eea
Here summation of $i,j$ indices over the three generations is understood.  
Among the 4 purely leptonic 
and 7 lepton-quark contact terms in the most general  
effective Lagrangian\cite{bchz97b}, only the above two terms with the SU(2) 
generators $T^a=\sigma^a/2$ contribute to the CC processes.    
We find 
\beq
\eta_{LL}^{\ell \nu_\ell \nu_{\ell'} \ell'} = {1\over 2}\,\eta_2^{LL}
\eeq
for the purely leptonic terms and 
\beq
\eta_{LL}^{\ell \nu_\ell u_i d_j} 
= V^{\rm SM}_{u_i d_j}\,{1\over 2}\,\eta_2^{LQ}
\eeq
for the lepton-quark interactions in terms of the quark mass eigenstates,   
where $V^{\rm SM}_{u_i d_j}$ are the standard CKM matrix elements.  

By inserting 
\bea
 V_{u d_j}^{\rm obs} &=& V_{u d_j}^{\sm} \Bigl[\,1 -{1\over 2}\, 
\frac{\eta_2^{LQ} -\eta_2^{LL}}{4\sqrt{2}G_F}\,\Bigr] 
\eea
into eq.~(\ref{unitarity_exp}) and using the unitarity 
relation of the SM, eq.~(\ref{unitarity_sm}), we obtain 
\bea
{1\over 2}\,[\eta_2^{LQ} -\eta_2^{LL}] = (0.116 \pm 0.069)\, \tev^{-2}\,.
\label{eta_constraint}
\eea
As expected, there is a 1.7$\sigma$ signal in 
the difference of the lepton-quark and purely leptonic 
CC contact interactions. 
If one assumes $\eta_2^{LL}=0$~\cite{altarelli}, the above result
(\ref{eta_constraint})
constrains $\eta_2^{LQ}$. 
On the other hand, if we allow $\eta_2^{LL}$ to vary 
freely under the constraint (\ref{delta_t}), 
we find 
\bea
{1\over 2}\,\eta_2^{LQ} 
= \epsilon\frac{4\pi}{(\Lambda_{CC, \epsilon}^{\ell q})^2} 
= (0.09^{+0.14}_{-0.09})\, \tev^{-2}\,.
  \label{eta2_constraint}
\eea
By studying the probability distributions in the region
of $\epsilon>0$ and $\epsilon<0$ separately, 
as in eqs.~(\ref{probabiblity1_norm}) and (\ref{probabiblity2_norm}), 
we obtain the following 95\%~CL bounds:
\bsub
\bea
\Lambda_{CC, +}^{\ell q} &>& 5.8\,\tev  \,, \\ 
\Lambda_{CC, -}^{\ell q} &>& 10.1\,\tev \,.
\eea
\esub
Here we allow $\eta_2^{LL}$ to vary freely, taking both
signs. 
If we assume no new physics contribution to the 
purely leptonic charged currents, $\eta_2^{LL}=0$, 
then the constraint (\ref{eta_constraint}) gives 
\begin{subequations}
\label{lambda_lq95}
\begin{eqnarray}
\Lambda_{CC, +}^{lq} &>& 7.4\,\tev \:\:\: ({\rm 95\%~CL})\,, \\ 
\Lambda_{CC, -}^{lq} &>& 12.4\,\tev \:\:\: ({\rm 95\%~CL})\,. 
\end{eqnarray}
\end{subequations}
The above constraints (\ref{lambda_lq95}) differ
significantly from the ones quoted in \cite{altarelli}, 
because we obtained the bounds in the $\epsilon>0$
and $\epsilon<0$ regions separately. 
In view of the above constraint (\ref{eta2_constraint}), 
deviation from the SM in the CC  
cross section at $Q^2 \sim 15,000\,\gev^2$ due to $V-A$ contact 
interactions should be at the level of at most one per cent. 

\section*{Acknowledgements}
The authors wish to thank K.\,Tokushuku for calling our attention to the 
charged-current data at HERA, and  
V.~Barger and D.~Zeppenfeld for enlightening discussions.  
They also wish to thank S.~Vascotto for careful reading of the manuscript.
This research was supported in part by the JSPS-NSF Joint Research Project.
The work of S.M. is supported by the CERN-Asia Fellowship. 
%


\newpage

\begin{figure}[t]
\begin{center}
\leavevmode\psfig{figure=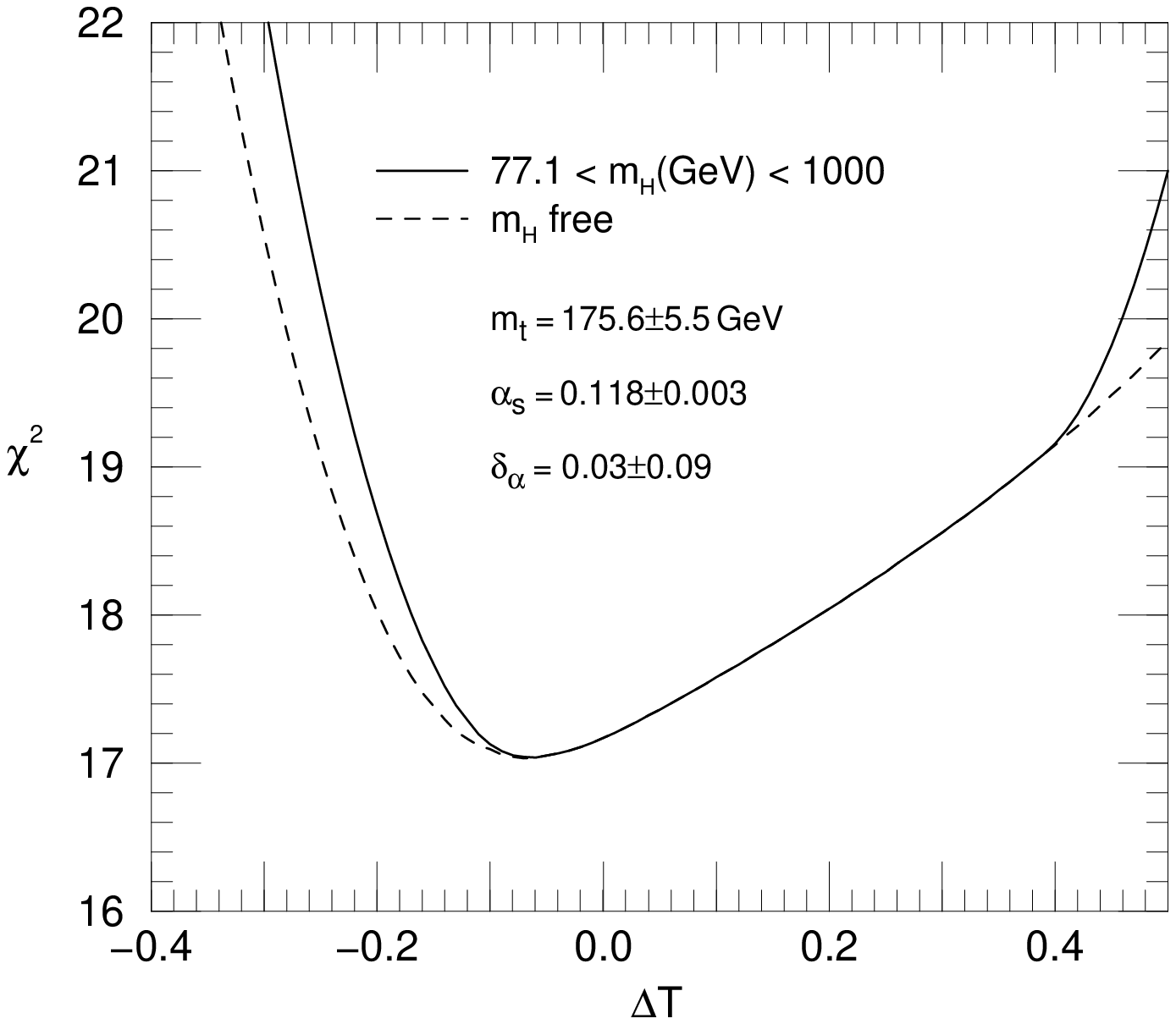,height=10cm}
\end{center}
\caption{
The minimal $\chi^2$ versus $\Delta T$ when $\mh$ is allowed to vary.  
The solid line is obtained by constraining $\mh$ in the 
range $77.1\, \gev <\mh<1\,\tev$ while the dashed line 
is obtained by allowing $\mh$ to vary freely. 
}
\end{figure}


\begin{thebibliography}{99}

\bibitem{h197} 
The H1 Collaboration, C.~Adloff et al., Z.~Phys.\ {\bf C74}, 191 (1997).

\bibitem{zeus97}
The ZEUS Collaboration, J.~Breitweg et al., Z.~Phys.\ {\bf C74}, 207 (1997).

\bibitem{altarelli97}
G.~Altarelli, J.~Ellis, G.F.~Giudice, S.~Lola, and M.L.~Mangano,
preprint CERN-TH/97-040 [hep-ph/9703276].

\bibitem{babu97}
K.S.~Babu, C.~Kolda, J.~March-Russell, and F.~Wilczek, Phys.\ Lett.\ {\bf
B402}, 367 (1997).

\bibitem{bchz97}
V.~Barger, K.~Cheung, K.~Hagiwara, and D.~Zeppenfeld,
Phys. Lett. {\bf B404}, 147 (1997). 

\bibitem{contact97}
A.~Nelson, Phys. Rev. Lett. {\bf 78}, 4159 (1997); 
N.~Bartolomeo and M.~Fabbrichesi, Phys.\ Lett.\ {\bf B406}, 237 (1997); 
M.C.~Gonzalez-Garcia and S.F.~Novaes, hep-ph/9703346;
W.~Buchm\"{u}ller and D.~Wyler, hep-ph/9704317; 
S.~Godfrey, Mod.\ Phys.\ Lett.\ {\bf A12}, 1859 (1997); 
N.G.~Deshpande, B.~Dutta, and Xiao-Gang~He, hep-ph/9705236; 
L.~Giusti and A.~Strumia, hep-ph/9706298;
Z.~Cao, X.-G.~He, and B.~McKellar, hep-ph/9707227.  

\bibitem{chm97}
G.-C. Cho, K. Hagiwara and S. Matsumoto, hep-ph/9707334.

\bibitem{bchz97b}
V. Barger, K. Cheung, K. Hagiwara, and D. Zeppenfeld, hep-ph/9707412.

\bibitem{elp82} 
E.~Eichten, K.~Lane, and M.~Peskin, Phys.\ Rev.\ Lett.\ {\bf 50}, 811 (1982).

\bibitem{pdg96}
Review of Particle Properties, Phys. Rev. {\bf D54}, 1 (1996).

\bibitem{heracc97}
B.~Straub, talk presented at Lepton-Photon '97 Symposium, Hamburg, 1997;\\  
E.~Elsen, International Europhysics Conference on High Energy Physics 
(EPS97) 19-26 August 1997, Jerusalem, Israel.

\bibitem{rueckl83}
R.~R\"uckl, Phys.\ Lett.\ {\bf B129}, 363 (1983); 
Nucl.\ Phys.\ {\bf B234}, 91 (1984).

\bibitem{burges84}
C.~J.~C.~Burges and H.~J.~Schnitzer, Phys. Lett. {\bf 134B}, 329 (1984).

\bibitem{cashmore85} 
R.J. Cashmore et al., Phys.\ Rev.\ {\bf 122}, 275 (1985).

\bibitem{burgess94}
C.P.~Burgess, S~Godfrey, H~Konig, D.~London, I.~Maksymyk, 
Phys.\ Rev. \ {\bf D49}, 6115 (1994).

\bibitem{chiappetta96} 
P.~Chiappetta and J.-M.~Virey, Phys.\ Lett.\ {\bf B389}, 89 (1996).

\bibitem{cornet97}
F. Cornet and J. Rico, Phys.\ Lett.\ {\bf B412}, 343 (1997). 

\bibitem{altarelli}
  G.~Altarelli, G.F.~Giudice, and M.L.~Mangano, 
  preprint CERN-TH/97-101 [hep-ph/9705287].

\bibitem{hhkm94}
K. Hagiwara, D. Haidt, C.S. Kim and S. Matsumoto,
Z. Phys. {\bf C64}, 559 (1994); Errata {\bf C68}, 352 (1995).

\bibitem{hhm97}
K. Hagiwara, D. Haidt and S. Matsumoto, hep-ph/9706331, 
to be published in Z. Phys. {\bf C} (1997).

\bibitem{stu}
M.E. Peskin and T. Takeuchi, Phys.\ Rev.\ Lett.\ {\bf 65}, 964 (1990); 
Phys.\ Rev.\ {\bf D46}, 381 (1992).

\bibitem{ew97}  
  D.~Ward, talk presented at
  International Europhysics Conference on High Energy Physics 
  (EPS97), 19-26 August 1997, Jerusalem, Israel.
%
\bibitem{mhlimit97}  
  P.~Janot, talk presented at
  International Europhysics Conference on High Energy Physics 
  (EPS97), 19-26 August 1997, Jerusalem, Israel.
%
\bibitem{eidelman95} 
S. Eidelman and F. Jegerlehner, Z.\ Phys.\ {\bf C67}, 585 (1995).

\bibitem{mz95}
A.D. Martin and D. Zeppenfeld, Phys.\ Lett.\ {\bf B345}, 558 (1995).

\end{thebibliography}
\end{document}